# A high-performance integrated single-photon detector for telecom wavelengths

DONALD S. BETHUNE, WILLIAM P. RISK AND GARY W. PABST[†]

IBM Almaden Research Center, 650 Harry Road, San Jose, CA 95120-6099
[†]Current address: Dept. of Electrical & Computer Eng., U.C. Davis, Davis, CA 95616-5294

**Abstract.** We have integrated a commercial avalanche photodiode (APD) and the circuitry needed to operate it as a single-photon detector (SPD) onto a single PC-board. At temperatures accessible with Peltier coolers (~200-240K), the PCB-SPD achieves high detection efficiency (DE) at 1308 and 1545 nm with low dark count probability (e.g. ~$10^{-6}$/bias pulse at DE=20%, 220 K), making it useful for quantum key distribution (QKD). The board generates fast bias pulses, cancels noise transients, amplifies the signals, and sends them to an on-board discriminator. A digital blanking circuit suppresses afterpulsing.

## 1. Introduction

The upsurge of interest in quantum key distribution (QKD) [1,2] in recent years has motivated an extensive effort aimed at developing single-photon detectors (SPDs) for the telecommunication wavelength windows around 1310 and 1550 nm [3-22]. The efficiency, dark count probability, and recovery times of the detectors limit the range and bit generation rates that can be achieved in QKD systems. Avalanche photodiodes (APDs) based on either Ge or InGaAs have proven to be the most convenient and cost-effective devices for this purpose. While the intrinsic characteristics of APDs are primary factors in determining their suitability for single-photon detection and have been intensively studied, to be useful as a SPD the photodiode must be surrounded with a constellation of auxiliary electronics. Single photons are detected using an APD by reverse-biasing the diode above its breakdown voltage, $V_{br}$. A single photoexcited carrier can then initiate an avalanche, which generates a large output charge pulse. In gated mode, the diode is DC biased slightly below $V_{br}$, and is pulsed above $V_{br}$ just at the photon arrival time to maximize the detection efficiency (DE). DE is the overall probability of registering a count if a photon arrives at the detector, and includes fiber coupling loss, APD optical coupling efficiency and intrinsic quantum efficiency, and the efficiency with which the signal processing electronics respond to photon signals from the APD. The bias pulse should be as short as possible to minimize the probability that a thermal carrier will be present during the bias pulse and will trigger an avalanche. Our PC-board single-photon detector (PCB-SPD) design generates 4.5 V, 1.2 ns bias pulses. Such pulses, applied to the APD cathode, generate extremely fast bipolar capacitive transients at the anode, since the reverse-biased EPM239aa APDs used in this work [23] have capacitance $\cong 0.25$ pF when reverse-biased near $V_{br}$ (~30-60 V). In earlier work on our prototype autocompensating QKD system [9,10], we introduced a circuit to eliminate these transients, so that the avalanche charge-pulse signal appears against a flat baseline. This signal is isolated from noise spikes by a fast electronic gate, amplified, and sent to a discriminator that gives a logic pulse out only if the signal pulse-height exceeds a set threshold. The circuit allows relatively small charge pulses ($10^5$-$10^6$ electrons) to be detected, increasing the DE. At the same time, the reduced average APD current reduces afterpulsing, which occurs when a charge trapped at a defect during an earlier avalanche event is released by a subsequent bias pulse and seeds a new avalanche.

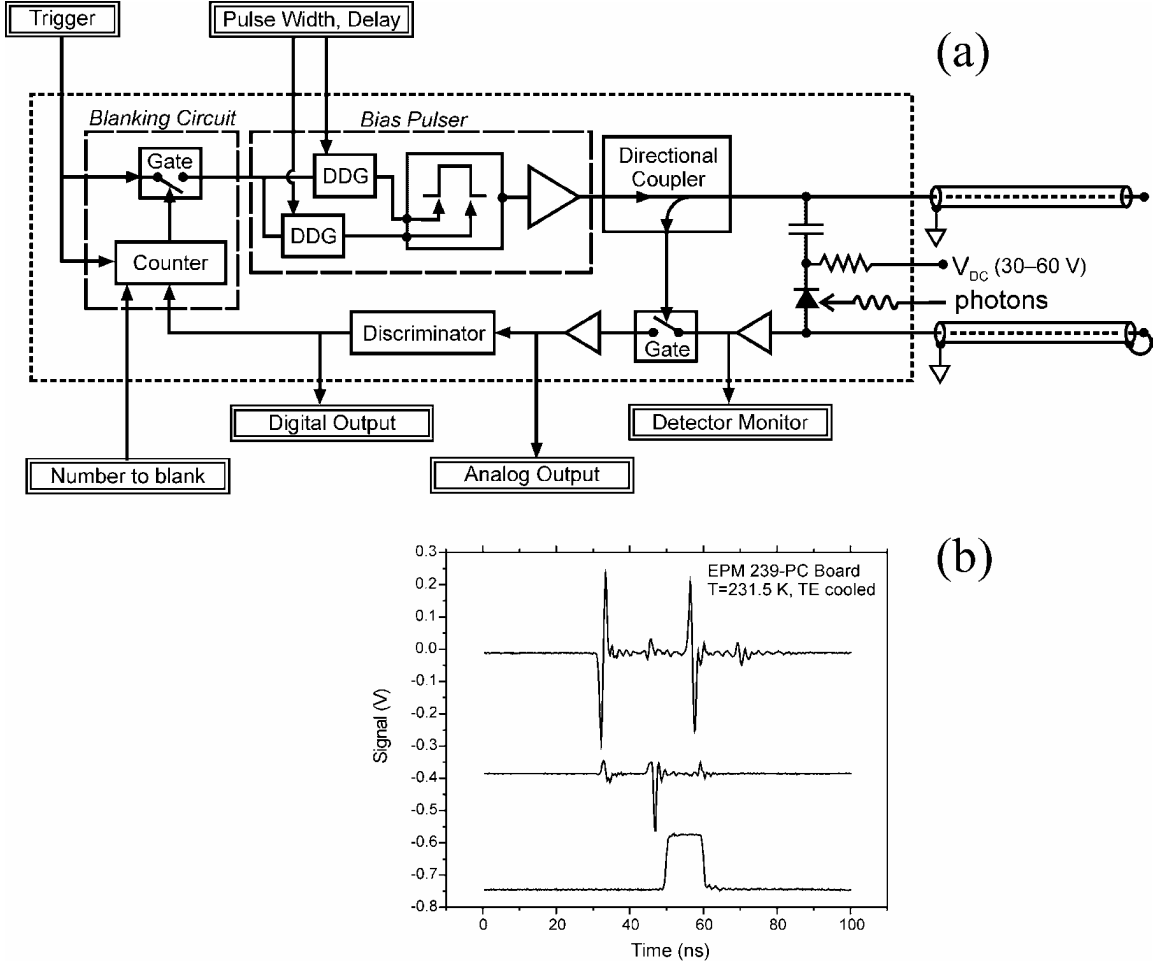

Figure 1. a) Schematic block diagram of PCB-SPD; b) scope traces showing *Detector Monitor* signal before gate (upper trace), *Analog Output* signal before the discriminator (middle trace), and *Digital Output* from discriminator (lower trace) for a pulse with an avalanche event.

In this paper we show that by integrating the auxiliary electronics required for single-photon detection onto a single PC board with the APD, we can obtain significantly improved overall single-photon detector performance. We attribute this improvement to the use of GHz-bandwidth surface-mount components in close proximity to the APD. These components generate 1 ns bias pulses with fast rise and fall times, and gate out a short slice of the APD output signal (also ~1 ns) to effectively isolate it from noise. In addition, a digital blanking circuit is used to turn off the bias pulses for a set number of triggers after a detection event, which greatly suppresses afterpulsing [15,16].

## 2. Experimental

Figure 1a shows a block diagram of the 95 x 190 x 1.5-mm$^3$ PCB-SPD. High-performance surface-mount emitter-coupled-logic (ECL) chips with 150-300-ps transition times are used throughout. A trigger pulse passes through the blanking circuit to a pair of digital delay generators (10-bit, with 10-ps resolution), which control the rising- and falling-edge positions of the bias pulse. The XOR of the delay outputs drives a 1-GHz bandwidth monolithic amplifier that gives a 4.5-volt pulse with width adjustable between ~0.9 and 5 ns. We obtained the best performance with a nominal 1-ns width (measured FWHM=1.2 ns). The pulse is sent through a 10dB directional coupler to the APD cathode, and then continues on 50 Ω stripline to an SMA edge connector where it is launched onto a 120-cm delay line. It reflects without inversion from the open end of the cable, and travels back toward the APD. The APD anode is coupled to a matched delay line, terminated with a short. A capacitive transient, sometimes with an avalanche signal pulse superimposed, travels on the lower delay line synchronously with the bias pulse on the upper line. It reflects, with inversion, from the shorted cable end, and arrives at the anode just as the reflected bias pulse arrives at the cathode. The transient generated by the reflected bias pulse cancels the inverted transient so that the (inverted) avalanche signal now appears on a flat baseline. The upper scope trace in Fig. 1b shows the amplified APD output (*Detector Monitor* in Fig. 1a), with the initial (transient + photon) signal, the net signal in the center of the trace, and finally the inverted transient due to the reflected bias pulse. Because of the highly linear nature of the tees and delay lines, the transients accurately cancel independent of bias pulse shape and of cable dispersion and attenuation. The reflected bias pulse serves one more function. A -10 dB (in terms of power, ~1.4 V in amplitude) portion of the reflected pulse is split off by a directional coupler and used to drive a back-to-back pair of surface-mount double-balanced mixers that serve as a fast gate. Since the reflected pulse on the upper delay line travels in parallel with the reflected signal on the lower delay line, the mixers pass only the 1-ns portion of the APD signal centered on the background-subtracted, photon-triggered avalanche pulse. The center trace in Fig. 1b (*Analog Output* in Fig. 1a) shows a photon signal after the gate and an additional (inverting) amplification stage. This signal is sent to a fast comparator with adjustable threshold that serves as a discriminator and outputs a logic pulse, as shown by the lower trace in Fig 1b (*Digital Output* in Fig. 1a).

The APD was embedded in an insulating foam cube (~5 x 5 x 5 $cm^3$), and could be cooled to temperatures, T, down to ~230 K using a pair of 2-stage thermoelectric coolers (TECs). By slightly cooling the TEC heatsink, the APD could be further cooled to 220 K. To determine whether it would be advantageous to go to still lower temperatures, we also used a test-setup with the APD thermally contacted to a copper tail that extended down into a liquid nitrogen dewar. In this case the APD temperature controller, which monitors a sensor attached to the APD holder, drove heaters on the Cu tail just below where it passed through the insulating foam into the APD chamber. This arrangement was used to reach temperatures down to 160 K.

To characterize the PCB-SPD, a 1308-nm distributed feedback laser (DFB) was pulsed at rates between 250 kHz and 2.5 MHz to give gain-switched ~50 ps pulses. Data were also obtained using a 1545-nm DFB laser. A variable attenuator and calibrated attenuation chain, which included two 95:5 splitters and fixed 5- and 10-dB fiber attenuators, attenuated the light pulses to give, on average, 0.1 photons/pulse at the APD fiber-pigtail coupler. An optical power meter monitored the 95% portion of the beam from the first splitter. The detector was biased at twice the light pulse rate, so that the APD was biased both in coincidence with the light pulses and midway between them. Counts for the laser-synchronized and interleaved bias pulses were accumulated by separate counters (L & D), allowing us to measure the dark count rate both with the light on and off. By comparing the light-on 'interleaved dark rate' measured by channel D to the light-off 'dark rate', we can quantify the afterpulse probability.

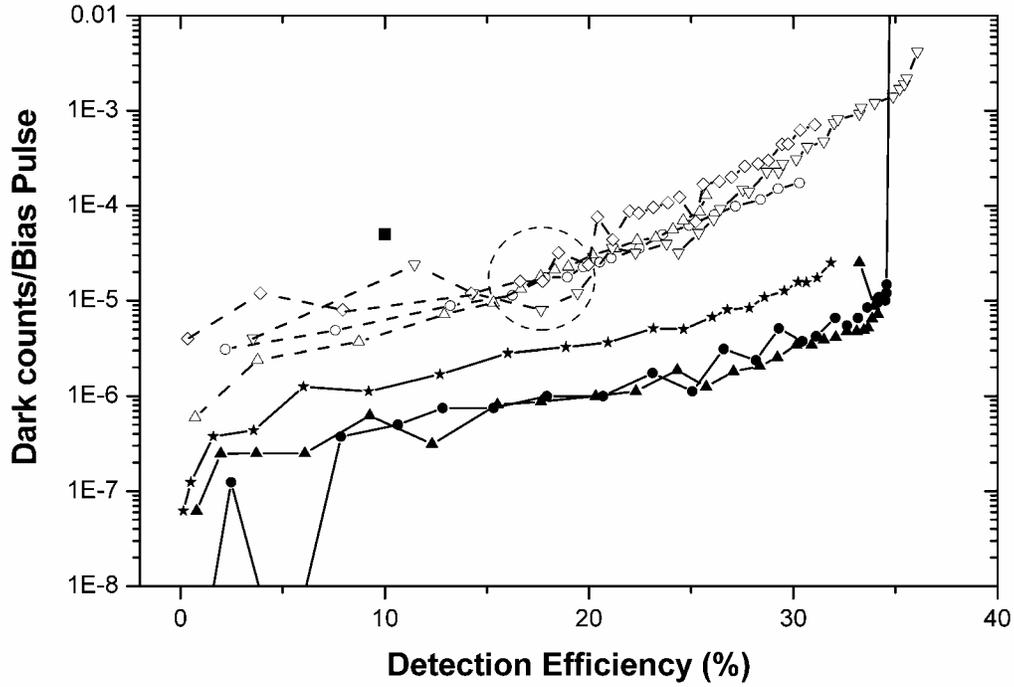

Figure 2. Dark count probability per bias pulse versus DE for: (●, ▲) 2 PCB-SPDs with EPM239 APDs at 220 K, 1308 nm light, bias frequency 500 kHz; (★) PCB-SPD with 239 K APD, 1545 nm light; (∇, ◊) a third EPM239aa, used with discrete-component bias circuit at 193 and 213 K; (o, ∆) 2 Fujitsu FPD5W1KS ADPs at 118 K. Also shown (■) is the specified performance for the id Quantique id200™ single-photon detector at ~220 K. The dashed circle indicates the typical operating conditions used previously with our prototype QKD system.

## 3. Results

A key performance measure for SPDs in applications such as QKD is the probability, $P_d$, of getting a dark count per bias pulse compared to the DE. Figure 2 shows $P_d$ plotted against DE for two PCB-SPDs with the APDs at 220 K and 1308 nm light, and for a PCB-SPD with 1545 nm light with the APD at 240 K. The performance achieved is compared to the best results we achieved earlier with Fujitsu FPD5W1KS InGaAs APDs [24] cooled to 118 K as used in our prototype QKD system, to the performance of another EPM239aa APD at 193 and 213 K with the same discrete-component biasing and detection circuitry used with the Fujitsu devices, and to the specified performance of the id Quantique id200™ EPM239-based single-photon detector [25]. The PCB-SPD dark count probability is significantly lower than any we have previously measured, including another EPM239aa at 193 K used with our discrete component bias circuitry, despite the relatively high operating temperature. At 220 K, $P_d \sim 10^{-6}$ at DE=20% with 1308-nm light, and at 240K, $P_d \sim 4 \times 10^{-6}$ at DE=20% for 1545-nm light. Between 240 and 273 K, the DEs measured for 1308 and 1545 nm light were essentially the same, with maximum DEs above 30%.

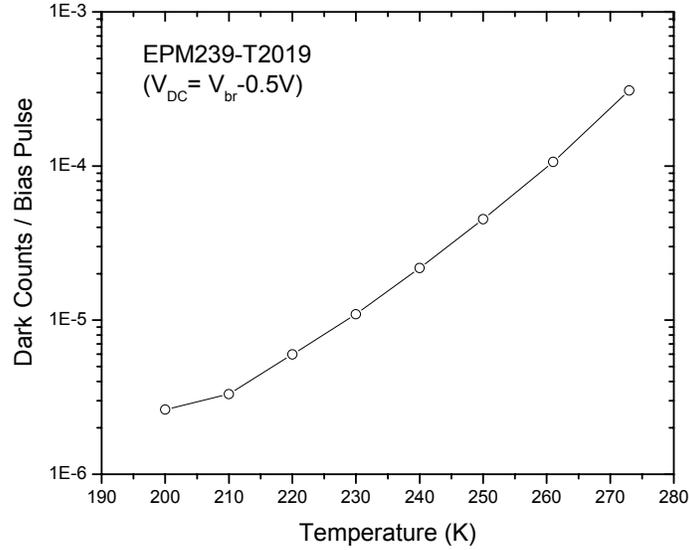

Figure 3. Dark counts per bias pulse ($P_d$) versus temperature for one PCB-SPD. The EPM239 APD was biased at 500 kHz. The points were measured with DC bias set 0.5 V below breakdown.

Much of the recent work on SPDs for telecom wavelengths has focused on the EPM239 APDs [12-22]. The $P_d$ versus DE values reported in these papers generally fall near the upper set of curves in Fig. 2. Recently Tomita, et al. have reported exceptionally low values of $P_d$ (~$2 \times 10^{-7}$ at DE=10%, 188 K, 1545 nm) [21-22]. This low-noise performance was achieved using an alternative capacitive-transient canceling scheme, which again allows small avalanche charge pulses to be detected [20]. In their arrangement a matched pair of EPM239ba APDs, driven by a common bias pulse, functions as a dual channel SPD. The outputs of the two APDs are electronically subtracted to cancel the common mode capacitive transient, leaving a positive- or negative-going avalanche pulse, depending on which APD fired. This elegantly simple and economical arrangement achieves dark count probabilities comparable to those obtained with our PCB-SPD, but at detector temperatures 20-30 K colder. This additional cooling is problematic for standard TECs. For QKD and other dual channel applications, the differential arrangement has two potential drawbacks. First, because the APDs share a single bias pulse, no *electronic* relative delay is possible, so to bring both optical pulses into coincidence with the bias pulse requires a fine *optical* delay adjustment in one channel. Second, in this arrangement the dual-SPD acts as an XOR gate, which gives no response for coincident photon detections. In some applications, including QKD, measuring the coincidence rate for the two channels may be desirable, since this rate provides additional information that is diagnostic for certain attacks. For example, if Eve determines the photon states for certain pulses, she could increase Bob's chance of detecting those pulses by replacing them with multiphoton pulses in the same states. But this would lead to an anomalously high coincidence rate that could alert Bob to the attack.

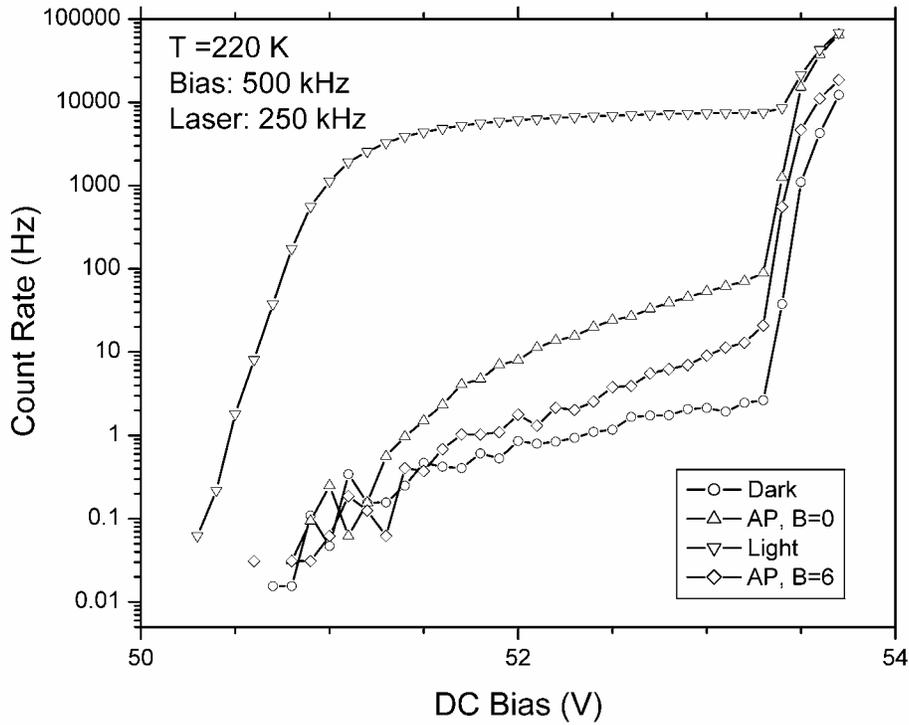

Figure 4.  **Light** (∇) and **dark** (o) count rates for PCB-SPD with light on (channel L, 0.1 photon/pulse) and off ([L+D]/2), respectively; **AP, B=0** (Δ) - the (afterpulse+dark) rate (channel D, 0.1 photon/pulse, no blanking); **AP, B=6** (◊) - (afterpulse+dark) rate (channel D, 0.1 photon/pulse, 6 bias pulses blanked).

Figure 3 shows the variation of $P_d$ with temperature for one PCB-SPD, with its DC bias set 0.5 V below $V_{br}$ for each temperature.  The dark counts fall by approximately a factor of 2 for each 10 K of cooling down to 200 K.  At 180 and 160 K, with a bias rate of 500 kHz, the dark count probability rose again.  We attribute this to increasing numbers of carriers in long-lived trap-states at low temperatures.  The release of such carriers during bias pulses gives rise to afterpulsing.  At 160 K this effect was so severe that runaway dark counts precluded normal operation of the PCB-SPD with a 500 kHz bias rate.  To characterize the afterpulse probability for the detector, we compared the dark count rate measured with the light off, to the interleaved dark rate measured with the light on at a standard intensity (0.1 photons/pulse at the APD pigtail).  A plot illustrating the effect is shown in Figure 4.  The upper and lower traces in Figure 4 (∇, o) correspond to the light-on counts for channel L and the light-off counts averaged for both channels, (L+D)/2.  (With the light off both channels are equivalent, and averaging reduces the statistical variation.)  The second curve (Δ) shows the light-on counts for the interleaved (D) channel.  The difference between this curve and the light-off dark rate is the afterpulse rate.  The afterpulse probability, $P_{AP}$, is defined as the ratio of this rate to the detector-firing rate.  We expect $P_{AP}$ to be proportional to the trapped charge density, which in turn will be proportional to the average APD current (and hence its firing rate). Various types of traps can have different (T-dependent) lifetimes.

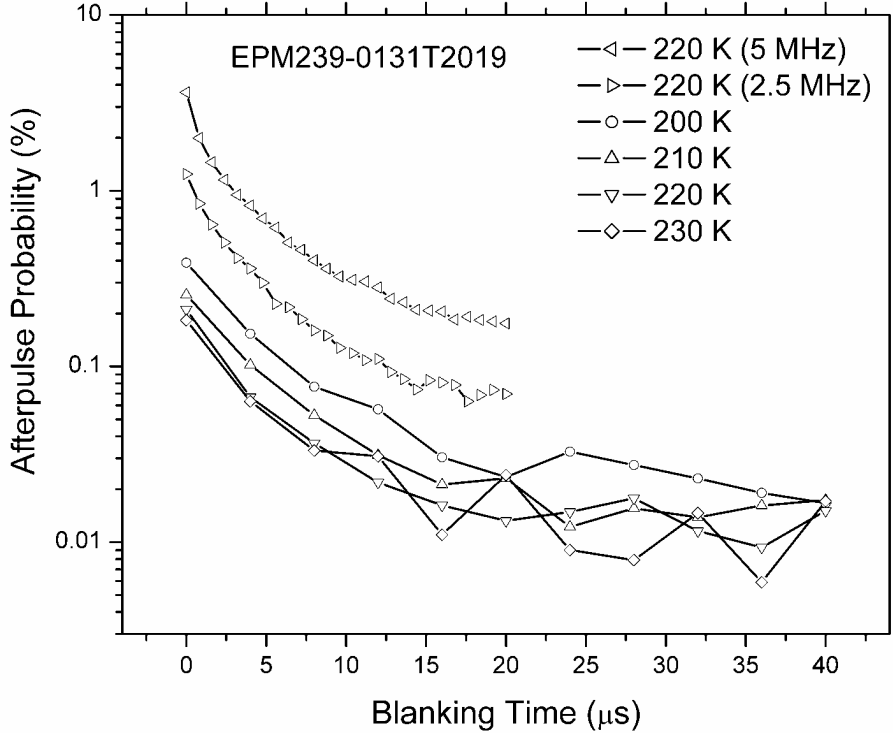

Figure 5. Afterpulse probability versus blanking time for various temperatures. The uppermost curves were measured at bias pulse rates of 5 and 2.5 MHz, while the others were measured at 500 kHz.

Afterpulsing can be partially suppressed by blanking the trigger pulses for an interval following each avalanche event [15,16]. We used a blanking circuit in which the discriminator output sets a flip-flop that enables an 8-bit trigger pulse counter, which in turn resets the flip-flop after reaching its preset count. The flip-flop also controls one input of an AND gate, allowing between 0 and 255 trigger pulses to be blocked. For our first PCB-SPDs, this blanking circuit was off-board (for the 2$^{nd}$ generation PCB-SPDs now being assembled it will be on-board). By measuring the afterpulse rate as a function of the blanking time, information about the decay of the trapped charges can be obtained. The results from such a set of measurements are shown in Figure 5, where $P_{AP}$ is plotted against the blanking time. The curvature of the plots on the semi-log graph indicates that a single exponential cannot describe the decays. However, the data show that $P_{AP}$ decays by more than 90% over a period of ~12 μs. Beyond that a long-lived component continues to contribute an afterpulse probability that does not decay within the measured time span. The third curve in Figure 4 (◊) shows an example of the afterpulse-rate reduction obtained by blanking 6 trigger pulses over 12 μs. At 52 V bias this rate is reduced by~85%. This reduction comes at the cost of a reduction in the effective DE by a dead-time factor $[1+ \mu\, DE\, N_B]^{-1}$, where μ is the mean number of photons/pulse and $N_B$ is the number of light pulses blanked after a detection event. For example, for $N_B$=3, μ = 0.1 and DE = 0.25 (with no blanking), the dead time correction factor is 0.93.

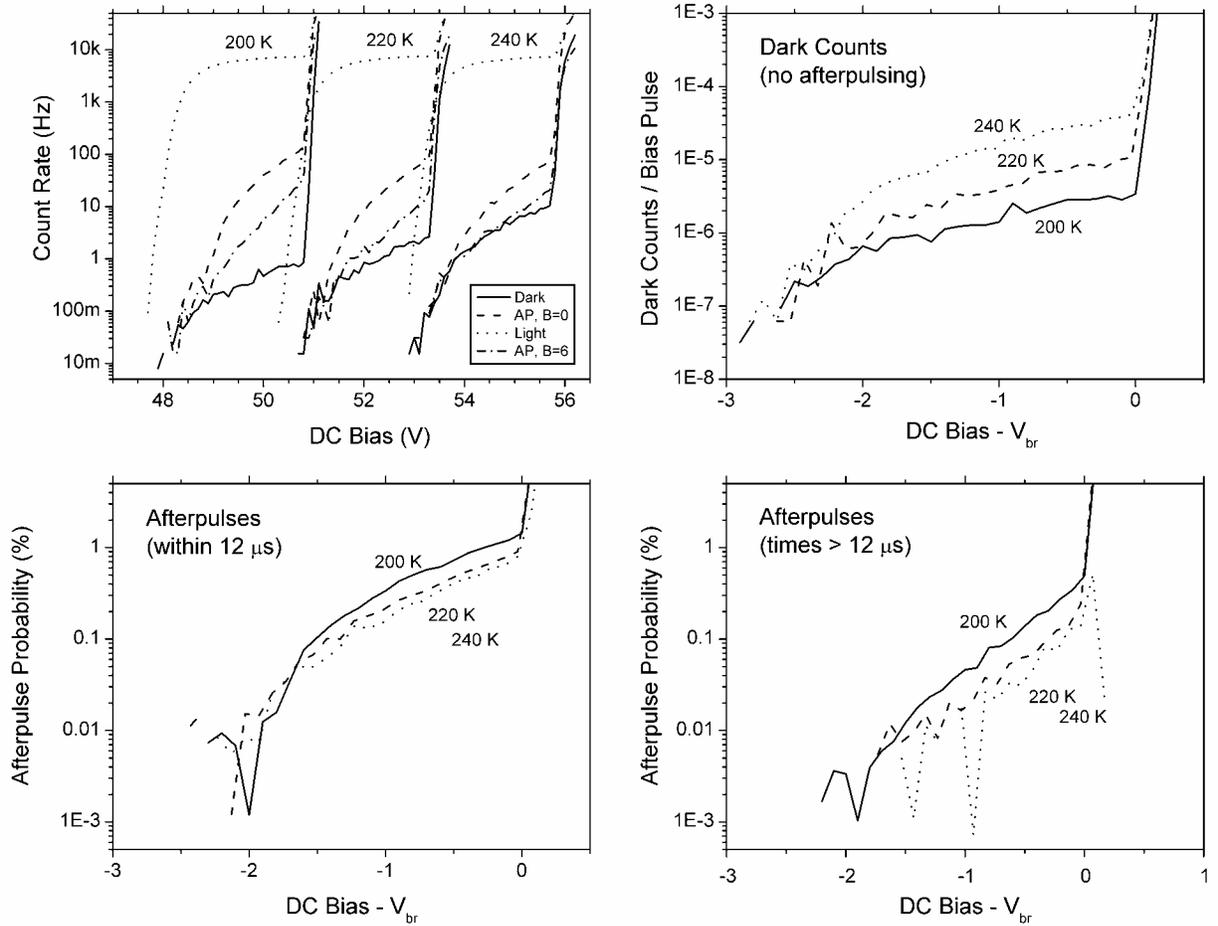

Figure 6. (Upper left) count rates as in Figure 4 for 200, 220 and 240 K; (upper right) light-off dark count probability, $P_d$ vs. DC bias voltage relative to $V_{br}$; (lower left and right) the portions of the afterpulse probability that decay in times ≤ 12 μs and >12 μs, respectively.

Figure 4 shows that at high detection rates (here ~7 kHz), afterpulsing can be the dominant source of noise counts for the PCB-SPD at low T, dominating the intrinsic dark count rate by more than an order of magnitude. Even with blanking, the long-lived component of $P_{AP}$ is still comparable to $P_D$. Since afterpulsing can be the primary limitation on SPD performance, we studied its temperature and bias voltage dependence in more detail. Representative data are presented in Figure 6. The upper left plot shows data similar to that in Figure 4 for APD temperatures 200, 220 and 240 K. With no blanking, the overall (dark+afterpulse) rate (dashed curves) actually rises slightly on cooling with the relatively high detection rates typical for these measurements, even though the dark count rate (solid curves and upper right panel) falls by an order of magnitude on cooling from 240 to 200 K. We note that the fraction of the afterpulsing rate due to long-lived traps increases rapidly as the temperature falls, limiting the effectiveness of blanking at low temperatures.

In the bottom two panels, the total afterpulse probability, $P_{AP}$, is split into two components: a long-lived component, $P_{APL}$, measured with 12 µs blanking (lower right), and a short-lived component, $P_{APS} = (P_{AP} − P_{APL})$ (lower left). Down to 200 K, both components increase moderately as the detector is cooled, but similar measurements at 180 K on a second APD suggest that $P_{APL}$ increases sharply below 200 K. $P_{APS}$ and $P_{APL}$ differ in their bias voltage dependences. Most strikingly, $P_{APL}$ increases exponentially with DC bias voltage at a rate of ~10x per volt. This behavior persists down to 180 K. This strong dependence of $P_{APL}$ on bias voltage, together with the ineffectiveness of blanking in suppressing this subset of afterpulses, will push the optimum operating point to lower bias values at high detection rates, thus limiting the useful detector efficiency.

## 4. Conclusions

The high performance of the PCB-SPD described in this paper highlights the advantages of integrating the primary photosensitive device and the electronics required to operate it as a single-photon detector. The high detection efficiency and low dark count probability this integrated design achieves will allow fiber-optic QKD over distances in the 100 km range, or higher bit generation rates over shorter distances. The PCB-SPD performed best with T in the range 200-220K. However, even at T=273 K the detector had a $P_d$ ~$3x10^{-5}$ with DE=20%, which would allow QKD systems to be operated over moderate distances with minimal detector cooling. This would greatly reduce the TEC power requirement.

The use of a digital blanking circuit enabled us to reduce the afterpulsing by about an order of magnitude by suppressing bias pulses for a time ~12 µs. However, we observe that afterpulsing due to carriers in long-lived traps increases with both increasing bias frequency and higher bias voltages, and is thus a barrier to increasing the bit generation rate for APD-based QKD systems.

**Acknowledgment**

This work was funded in part by the DARPA-QuIST program.